\def \SAIT #1 #2 {{\em Mem.\ Soc.\ Astron.\ It.\/} {\bf #1}, #2}
\def \MESS #1 #2 {{\em The Messenger\/} {\bf #1}, #2}
\def \ASTRNACH #1 #2 {{\em Astron. Nach.\/} {\bf #1}, #2}
\def \AAP #1 #2 {{\em Astron. Astrophys.\/} {\bf #1}, #2}
\def \AAL #1 #2 {{\em Astron. Astrophys. Lett.\/} {\bf #1}, L#2}
\def \AAR #1 #2 {{\em Astron. Astrophys. Rev.\/} {\bf #1}, #2}
\def \AAS #1 #2 {{\em Astron. Astrophys. Suppl. Ser.\/} {\bf #1}, #2}
\def \AJ #1 #2 {{\em Astron. J.\/} {\bf #1}, #2}
\def \ANNREV #1 #2 {{\em Ann. Rev. Astron. Astrophys.\/} {\bf #1}, #2}
\def \APJ #1 #2 {{\em Astrophys. J.\/} {\bf #1}, #2}
\def \APJL #1 #2 {{\em Astrophys. J. Lett.\/} {\bf #1}, L#2}
\def \APJS #1 #2 {{\em Astrophys. J. Suppl.\/} {\bf #1}, #2}
\def \APSS #1 #2 {{\em Astrophys. Space Sci.\/} {\bf #1}, #2}
\def \ASR #1 #2 {{\em Adv. Space Res.\/} {\bf #1}, #2}
\def \BAIC #1 #2 {{\em Bull. Astron. Inst. Czechosl.\/} {\bf #1}, #2}
\def \JSQRT #1 #2 {{\em J. Quant. Spectrosc. Radiat. Transfer\/} {\bf #1}, #2}
\def \MN #1 #2 {{\em Mon. Not. R. Astr. Soc.\/} {\bf #1}, #2}
\def \MEM #1 #2 {{\em Mem. R. Astr. Soc.\/} {\bf #1}, #2}
\def \PLR #1 #2 {{\em Phys. Lett. Rev.\/} {\bf #1}, #2}
\def \PASJ #1 #2 {{\em Publ. Astron. Soc. Japan\/} {\bf #1}, #2}
\def \PASP #1 #2 {{\em Publ. Astr. Soc. Pacific\/} {\bf #1}, #2}
\def \NAT #1 #2 {{\em Nature\/} {\bf #1}, #2}

\documentstyle{memsait}
\input epsf.sty
\begin{opening}
\title{BLANK FIELD X-RAY SOURCES} % ALL CAPITAL LETTERS PLEASE !!!
\author{Ilaria Cagnoni$^1$, Annalisa Celotti$^1$, Martin Elvis$^2$, Dong-Woo
Kim$^2$ $^3$,
Fabrizio Nicastro$^2$
}
\institute{$^1$International School for Advanced Studies (SISSA-ISAS), Trieste, Italy\\
$^2$Harvard-Smithsonian Center for Astrophysics, Cambrdige, MA, USA\\
$^3$Chungnam National Univ., Taejon, S.Korea}
\date{} % DO NOT INSERT ANY DATE HERE !!!
\end{opening}

\begin{document}

%\oddpagefooter{\sf Mem. S.A.It., Vol. ??, ??}{}{\thepage}
%\evenpagefooter{\thepage}{}{\sf Mem. S.A.It., Vol. ??, ??}
\oddpagefooter{}{}{} % LEAVE AS IT IS !
\evenpagefooter{}{}{} % LEAVE AS IT IS !
\ 
\bigskip

\begin{abstract}
The X-ray sky is not as well known as is sometimes thought. We report on our
search of minority populations (Kim \& Elvis 1999). One of the most intriguing
is that of `blank field sources', i.e. bright ROSAT sources ($F_X > 10^{-13}$ erg
cm$^{-2}$ s$^{-1}$) with no optical counterpart on the Palomar Sky Survey (to
O=21.5) within their $39^{\prime \prime}$ (99\%) radius error circle. \\
The nature of Blank Field sources is unknown: no known extragalactic population
has such extreme X-ray to optical ratios ($f_X/f_V >60$). 
Moreover blank field source tend to have much flatter PSPC spectra
compared to radio-quiet AGN. Both properties suggest obscuration. \\
`Blank field sources' could be: Quasar-2s, low-mass AGNs, isolated neutron
stars, high redshift clusters of galaxies, failed clusters, AGNs with no big
blue bump. Identification with any  of these populations would be an interesting 
discovery.
\end{abstract}

\section{Sample Selection}

\noindent Blank Field Sources (BFS) are extreme X-ray loud objects; 
the non detection of an optical counterpart down to the Palomar limit 
(O=21.5) imply  $f_X/f_V> 60$.

To isolate ROSAT BFS we selected the ROSAT PSPC pointed observation
sources with $f_x > 10^{-13}$erg~cm$^{-2}$s$^{-1}$ and with no optical
counterparts in the digitized sky surveys. 
We used all the sources from the WGACAT (White, Giommi \&
Angelini 1994) that were: (1) well-detected; (2) at high 
Galactic latitude ($|b|>$20$^{\circ}$);
(3) lay within the `inner circle' of the PSPC 
(for their smaller positional uncertainty).
We then used the APM digitization of the Palomar and UK Schmidt
sky surveys (McMahon \& Irwin, in preparation), via the WWW, to
find sources with no optical counterpart within the 99\% X-ray
position circle ($r\sim 39^{\prime\prime}$).
Due to a problem with rev~0 images used for WGACAT
 (Giommi, Angelini \& Cagnoni, private communication 1999) 
$\sim 35$\% of the BFS had wrong coordinates.
We quote here the correct positions based on rev~2 version 
(from the Galpipe catalog) of ROSAT data.
The total sample consists of 16 sources.

\section{Nature of Blank field Sources}

BFS nature is still mysterious:
no known extragalactic population, including normal quasars, AGNs,
 BL Lacertae objects, normal galaxies and clusters of galaxies,
can reach such an extreme $f_X/f_V$ (see Maccacaro
et al. 1988's nomograph).\\
The unsually large $f_X/f_V$ ratio, however, is not the only strong
peculiarity of this class of sources. X-ray colors reveal other
interesting properties:
% Figure 1 shows the color-color diagram of the
%16 BFS of our sample. The color distribution of
%radio-quiet AGNs is overimposed for comparison (small points without errorbars,
%Fiore et al., 1998).  
BFS tend to have flatter PSPC spectra, compared to radio-quiet AGNs, in
both the soft (0.1-0.8 keV) and the hard (0.9-2.4 keV) PSPC range.\\
No known source has 
such an extreme hard distribution in the $\alpha_S$-$\alpha_H$ plane 
(Fiore et al., 1998, Nicastro, Elvis \& Fiore 1998):
 Galactic X-ray 
sources have usually  very steep PSPC spectra, which in term of PSPC 
colors means $\alpha_S$ and $\alpha_H$ greater than 2; normal galaxies 
and/or cluster of galaxies have thermal spectra with typical temperature 
of 1 keV and are generally distributed within a region of flat 
$\alpha_S$ and quite steep $\alpha_H$; radio-quiet and 
radio-loud type 1 AGN are well described by a single power 
law in the PSPC range and so lie in the central region of the 
$\alpha_S-\alpha_H$ diagram; finally Seyfert 2, with their 
heavily absorbed intrinsic X-ray continua, have $\alpha_S$ and 
$\alpha_H$ typical of normal galaxies (the contribution from the host 
galaxy emerging in the soft X-ray band).  \\

%\begin{figure}
%\epsfysize=6cm % fix the y-dimension and scales x-dim. to y-dim.
%%\epsfxsize=8cm % fix the x-dimension and scales y-dim. to x-dim.
%% Feel free to do the choice you prefer but do not exceed the x-dimension
%% of the text lines
%\hspace{3.5cm}\epsfbox{fig1.ps} %for centering: act on hspace argument 
%\caption[h]{ Color-color diagram of the BFS of our 
%sample. The PSPC colors of the radio-quiet AGN are overimposed for 
%comparison (small points without errorbars)}
%\end{figure}

\noindent The still open possibilities regarding the nature of BFS are:\\
1) - Quasar~2s: high luminosity, high redshift, heavily obscured quasars, 
the bright analogs of the well known Seyfert 2s;\\
2) - low mass Seyfert~2: that is AGNs powered by a low mass obscured black hole
(i.e. obscured Narrow Line Seyfert 1);\\
3) - high redshift clusters of galaxies;\\
4) - failed clusters: (Tucker et al. 1995) in which a large overdensity of matter has collapsed but 
has not formed galaxies;\\
5) - AGNs with no big blue bump.\\
The identification of BFS with any of these would be an important find.\\
The faintness of their optical counterparts and the flatness of their PSPC
 spectra can be, in cases (1) and (2), due to the effect of the intrinsic 
absorption.

For all the BFS we obtained optical imaging in R and K
to R=22 (this corresponds
to B$\sim$23.5 for a galaxy) and K=20 and 5 out of 16 sources show red
(R-K$>4$) counterparts as expected from obscured AGNs and high z clusters of galaxies.
 We are applying to IR telescopes to obtain spectra of these 5 BFS with an interesting 
counterpart.
The other 11 BFS do not show any obvious counterpart and their X-ray error circles
 show some sources ($\sim$ 5) expected as chance coincidence at these optical/IR
 magnitude limits.

\acknowledgements
This work was supported by NASA grants NAG5-3066 and NAG5-9206(ADP) and the Italian
MURST (IC and AC).

\end{document}